\documentclass[draft,jgrga]{agutex}		

\usepackage[pdftex]{graphicx}  	

\setkeys{Gin}{draft=false}			

\authorrunninghead{MURSULA ET AL.}

\titlerunninghead{HSS OCCURRENCE OVER GMM}

\begin{document}


\title{Occurrence of high-speed solar wind streams\\ over the Grand Modern Maximum}


\authors{K. Mursula\altaffilmark{1}, R. Lukianova\altaffilmark{2,}\altaffilmark{3}, L. Holappa\altaffilmark{1} } 

%
%
%

\altaffiltext{1}{ReSoLVE Centre of Excellence, Department of Physics, University of Oulu, Finland}
\altaffiltext{2}{Geophysical Center of Russian Academy  of Science, Moscow, Russia}
\altaffiltext{3}{Space Research Institute, Moscow, Russia}

\begin{abstract} 

In the declining phase of the solar cycle, when the new-polarity fields of the solar poles are strengthened by the transport of same-signed magnetic flux from lower latitudes, the polar coronal holes expand and form non-axisymmetric extensions toward the solar equator.
These extensions enhance the occurrence of high-speed solar wind streams  (HSS) and related co-rotating interaction regions in the low-latitude heliosphere,
 and cause moderate, recurrent geomagnetic activity in the near-Earth space.
Here, using a novel definition of geomagnetic activity at high (polar cap) latitudes and the longest record of magnetic observations at a polar cap station, we calculate the annually averaged solar wind speeds as proxies for the effective annual occurrence of HSS over the whole Grand Modern Maximum (GMM) from 1920s onwards.
We find that a period of high annual speeds (frequent occurrence of HSS) occurs in the declining phase of each solar cycle 16-23.
For most cycles the HSS  activity clearly maximizes during one year, suggesting that typically only one strong activation leading to a coronal hole extension is responsible for the HSS maximum.
We find that the most persistent HSS activity occurred in the declining phase of solar cycle 18.
This suggests that cycle 19, which marks the sunspot maximum period of the GMM, was preceded by exceptionally strong polar fields during the previous sunspot minimum.
This gives interesting support for the validity of solar dynamo theory during this dramatic period of solar magnetism.

\end{abstract}

\begin{article}

\section{Introduction}

Properties of the solar wind (SW), the tenuous plasma emanating from the solar corona, have been directly monitored since the early 1960s, soon after the start of the space age. 
Figure \ref{figure1_bw}a shows the annual means of the solar wind speed at the Earth's orbit in 1964-2010, measured by a number of Earth-orbiting satellites and interplanetary probes collected in the NASA/NSSDC OMNI-2 database. 
One can see that solar wind speed depicts a cyclic variation, with maxima in the declining phase of the solar cycle (SC), when the polar coronal holes extend equatorwards and emit high-speed solar wind streams (HSS) at sufficiently low latitudes to be detected at the Earth \citep{Krieger_etal_73, Bame_1976, Kojima-Kakinuma-90, Rickett-Coles-91, Mursula_Zieger_1996}. 

It is known that the Sun experienced a period of exceptionally high sunspot activity during the 20th century \citep{Solanki_2004, Abreu_2008}, now termed the Grand Modern Maximum of solar activity. 
Recent observations of declining sunspot magnetic field strengths \citep{Penn_2006, Livingston_2009}, vanishing of small sunspots \citep{Lefevre_Clette_2011} and the weakness of the ongoing cycle 24 suggest that the GMM has reached its end. 
Because the maximum of the GMM was already reached during cycle 19 in the late 1950s, direct solar wind observations only exist from the declining phase of the GMM. 
Moreover, since systematic measurements of the solar global magnetic field started only in the 1970s, no direct information on solar polar fields exists from the ascending phase or the maximum of the GMM. 

Geomagnetic activity (GA), i.e., the short-term variability of the Earth's magnetic field, is driven by the solar wind and the interplanetary magnetic field (IMF). 
Figure \ref{figure1_bw}b depicts the annual means of one of the most commonly used measures of geomagnetic activity, the aa index, since 1920s \citep{Mayaud_1972}. 
The aa index depicts a somewhat different solar cycle variation than the SW speed. 
For cycles 20 and 23, the aa index and SW speed have maxima in the same years, but for cycles 21 and 22 the cycle maximum of the aa index is earlier. 
Also, the relative heights of the four solar cycle maxima are different in GA and in SW speed. 
These differences between solar wind speed and the aa index are due to the fact that not only high-speed streams but also coronal mass ejections (CME) are important SW structures driving geomagnetic activity at mid-latitudes, which is monitored by the aa index. 
However, since HSSs and CMEs have different solar cycle occurrences, they have different relative contributions to geomagnetic activity over the solar cycle \citep{Richardson_atal_2002a, Holappa_Mursula_2014}. 
While high-speed streams maximize in the declining phase, the CMEs follow the sunspot cycle fairly reliably and are therefore more common around sunspot maxima.

The dependence of geomagnetic activity on solar wind speed is most direct in the auroral region \citep{Finch_etal_2008}, where strong horizontal currents called the auroral electrojets flow in the ionosphere, the ionized layer of the upper atmosphere. 
During magnetospheric substorms, whose occurrence is maximized in the years of persistent high-speed solar wind streams \citep{Tanskanen_etal_2005}, the westward auroral electrojet (WEJ) is extended in local time and enhanced in intensity, especially in the evening to pre-midnight sector, causing a depletion in the horizontal magnetic field at stations close to the auroral oval. 
(This depletion and, thus, the strength of the WEJ is measured by the geomagnetic AL index). 
In addition to depleting the horizontal field, the WEJ increases the magnetic field poleward of the auroral oval in the polar cap, where the field is nearly vertical, i.e., almost purely aligned in the Z-direction.
Note that the WEJ increases the field in both hemispheres, since in the north (south, respectively) the main field is oriented into (away from) the Earth, and the WEJ produces a downward (upward) oriented field in the northern (southern) polar cap.

\section{HSS effect upon the Z-component in the polar cap}

Presenting a new way of studying the solar wind-magnetosphere coupling, it was recently shown  \citep{Lukianova_Mursula_2012} that the annual means of the magnetic field Z-component at polar cap observatories are significantly enhanced during the strongest HSS years. 
Figure \ref{figure2_bw} depicts the annual means in 1926-2009 (data gap in 2007-2008) of the Z-component at Godhavn (GDH; 69.25$^{\circ}$ geographic latitude, 306.47$^{\circ}$ geographic longitude), the longest running polar cap observatory. 
In addition to the full year means, Figure \ref{figure2_bw} also depicts the annual means of the Z-component in magnetically quietest days (called Z(q), calculated from the five quietest days of each month), as well in the most disturbed days (called Z(d), obtained from the five most disturbed days each month). 
(In 1932-2009 the internationally selected five quietest and five most disturbed days of each month are based on the geomagnetic Kp index, which exists from 1932 onward. 
For the earlier years 1926-1931 we have used these days based on the Ci index.
Magnetic data were received from the World Data Center at Edinburgh (http://www.wdc.bgs.ac.uk), and  solar wind data from NASA/NSSDC OMNI2 database (http://omniweb.gsfc.nasa.gov.)

The long-term (secular) evolution of the Z-component at GDH during the time depicted in Figure \ref{figure2_bw} is roughly sinusoidal with amplitude of about one percent of the field intensity. 
On top of the secular variation Figure \ref{figure2_bw} shows, particularly clearly in the Z(d) curve, several positive deflections, i.e., years of enhanced vertical magnetic field intensity. 
The two largest positive deflections occurred in 1952 and 2003, some others, e.g., in 1930, 1941, 1963, 1982 and 1994. 
It was shown recently \citep{Lukianova_Mursula_2012}, using a number of observatories from both the northern and southern polar cap, that the deflections in the Z-component occurred at all stations in the same years, when high-speed stream occurrence and substorm activity (and the related AL index) were maximized. 
Accordingly, the Z-component of a polar cap station can be reliably used to identify years of persistent HSS activity. 
The two small inserts in Figure \ref{figure2_bw} depict an enlarged view of the Z(d) and Z(q) curves around 1952 and 2003, the two years of the largest deflections. 
The quiet-day curve Z(q) is also somewhat increased in these years, indicating that the persistent high-speed streams raised the Z-component even during the quietest days above its normal level. 
In order to quantify the effect of high-speed streams on the Z-component, we remove the quiet time field Z(q) from the disturbed time field Z(d), constructing annual differences $\Delta Z$ = Z(d) - Z(q). 
This difference also removes the seasonal and secular variations of the absolute level of the magnetic field, which is necessary in view of the rather large related variations of the magnetic baseline.

\section{Annual solar wind speed from $\Delta Z$}

Figure \ref{figure3_bw}a depicts the annual means of $\Delta Z$ for GDH in 1926-2009. 
The cycle maxima of $\Delta Z$ are found in the declining phase of all the included 8 solar cycles, in agreement with the above discussed solar cycle occurrence of equatorial extensions of coronal holes and high-speed streams at the Earth's orbit. 
In fact, $\Delta Z$ has cycle maxima in the same years as SW speed (see Fig. \ref{figure1_bw}a) for all other cycles except for cycle 21, where SW speed attained closely similar values in a few consecutive years (1982-1985). 
Accordingly, the annual means of GDH $\Delta Z$ and the solar wind speed must be highly correlated.
(Figure \ref{figure3_bw}b will be discussed later in Section \ref{HSS_SOD}.)

The  scatter plot between the annual values of GDH $\Delta Z$ and the solar wind speed is depicted in Figure \ref{figure4_bw}a. 
(We note that it is more correct to present  $\Delta Z$ in terms of SW speed, since SW speed is the driver of  $\Delta Z$ and known very accurately while $\Delta Z$ is known much less accurately. 
However, practically the order of correlation makes little difference to the main results.)
The correlation coefficient between the annual means of GDH $\Delta Z$ and SW speed is r = 0.78 which, for 43 annual data points, yields the zero correlation probability of p = $5.6\cdot 10^{-5}$ using Student's t-test and an autoregressive (AR-1) noise model.
Thus, SW speed describes the long-term evolution of GDH $\Delta Z$ during the last four cycles quite well, explaining 60\% (r$^2$ = 0.60) of its inter-annual variation. 
(Thus, the other possible factors, in particular the IMF strength and orientation, have considerably less influence on GDH $\Delta Z$). 
Note that the three years (1974, 1994, 2003) of the highest speeds in cycles 20, 21, and 23 stand out in Fig. \ref{figure4_bw}a as the only years when the annual mean speed is larger than 500\,km/s.
Note also that the correlation is fairly linear, indicating applicability over the whole range of $\Delta Z$ and SW speed values.
(In fact, using a separate fit for high-speed years with, e.g., V$>$450\,km/s would not lead to greatly different results).
  
The corresponding annual SW speed values in 1926-2009 (except for 2007-2008) estimated from this correlation are shown in Figure \ref{figure4_bw}b, together with the observed SW speed values in 1965-2009.
We have also included there the 1$\sigma$ (68\,\%) errors for the SW proxies estimated in the following way.
We first calculated the standard deviation $\delta{_Z}$ of the residuals of the best fit solution depicted in Fig. \ref{figure4_bw}a.
Then we generated a new $\Delta Z$ series by adding a random error (following Gaussian distribution with zero mean and standard deviation $\delta{_Z}$) to each annual $\Delta Z$ value, and calculated the SW speed proxy for this series in the same way as for the original $\Delta Z$ series.
This was repeated 10.000 times, producing a distribution of annual values for each year, whose standard deviations were used as errors for the respective years.

One can see in Fig. \ref{figure4_bw}b that, overall, the GDH $\Delta Z$ values yield a very good proxy for the solar wind speed at annual resolution for the whole 45-year time interval. 
As already seen in Fig. \ref{figure4_bw}a, the proxy covers the range of SW speed values very well, reaching the highest peaks and the lowest minima of SW speed. 
There are very few years when the proxy deviates from the measured SW speed value by more than one standard deviation.
This occurs more systematically only in the ascending phase of cycle 21 (in 1977-1979), when the estimated speed is too large, and during the maximum years of cycles 21 and 23, when the estimates are low.

The annual SW speeds estimated from GDH $\Delta Z$ values (see Fig. \ref{figure4_bw}b) almost double the time of the directly measured solar wind, covering the whole period of high sunspot cycles of the Grand Modern Maximum. 
Figure \ref{figure4_bw}b shows the same long-term evolution as Fig. \ref{figure3_bw}a, now quantified as SW speed.
Even if the estimated errors are rather large (due to neglect of other, less important drivers and CMEs), one can find significant differences between the individual years and solar cycles during the early part of the time series, i.e., during the rising phase of the GMM.
Figure \ref{figure4_bw}b shows high SW speeds during three years of the declining phase of cycle 18, with maximum speeds in 1952.
Accordingly, we find here that the year 1952 was the most HSS active year during the whole GMM. 

The annual solar wind speeds during cycle 17 remained at rather similar, moderate level in a few consecutive years, thus reminding the situation during cycle 22.
However, during cycles 16 and 19 there was one year (1930 and 1963, respectively) with considerably higher HSS activity than in any other year of those cycles.
This reminds the situation during most of the recent cycles (cycles 20, 22 and 23), where high HSS activity  was limited to one year only. 
Note also that, overall, the HSS activity remained at a rather moderate level during cycle 19, even though this cycle was exceptional in sunspot activity.
Cycle 19 was very active also in the aa index (see Fig. \ref{figure1_bw}b), which attained larger values during this cycle (with maximum in 1960) than in the earlier cycles.
The high peak of the aa index in cycle 19 must therefore be mainly produced by CMEs, whose amount  increases with increasing sunspot activity during the rise of the GMM.
This comparison demonstrates the different effects of the two SW structures, with HSSs dominating the high-latitude GDH $\Delta Z$ and CMEs the mid-latitude aa index (see also \citet{Holappa_Mursula_2014}).


\section{Local time distribution and secular change of $\Delta Z$}

It is interesting to study how the auroral electrojets are seen in the mean daily distribution of $\Delta Z$ values and how the current systems evolve in time over the GMM period.
Figure \ref{figure5} shows the annual averages of $\Delta Z$ in color scale, separately for each hour of the day (vertical axis) and each year (horizontal axis).
Each narrow vertical color strip then presents the yearly averaged local time (LT) distribution of $\Delta Z$. 
Positive values (yellow to red in the color code of Fig. \ref{figure5}) depict the average LT location of the westward auroral electrojet and negative values (blue to black color) the eastward electrojet (EEJ).  
One can see in Fig. \ref{figure5} that the WEJ region is centered around the local midnight to post-midnight hours and the EEJ in the local noon to afternoon sector.
The WEJ region is wider in local time, about 12 hours on an average, while the typical extent of the EEJ region is about 8 hours.
Therefore, e.g., the total annual $\Delta Z$ values (used above), which are averages of the 24 hourly values depicted in Fig. \ref{figure5}, are all positive.

Figure \ref{figure5} shows that, in certain years, the WEJ extends over a wider LT range than on an average.
This is particularly clearly seen
as the extension of the WEJ region from the midnight (top of Fig. \ref{figure5}) toward the evening sector (downwards in Fig. \ref{figure5}).
(A corresponding extension also takes place in the morning, but generally remains less notable).
As seen in Fig. \ref{figure5} large extensions occur roughly once every solar cycle and, in fact, correspond exactly to the years of the most persistent HSS activity discussed above. 
For example, in the early 1950s the WEJ region extends from midnight until about 16 LT, reducing the average intensity and extent of the eastward electrojet in these years. 
The extension of the WEJ from the midnight toward the afternoon is related to the dynamics of the magnetosphere during substorms (in particular to the development of the westward traveling surge;
see, e.g.,  \citet{Lockwood_living_rev_2013}) 
whose occurrence is maximized during HSS years \citep{Tanskanen_etal_2005, Tanskanen_etal_2011}.
Accordingly, the connection between high-speed streams and the extensions of the WEJ are well understood.

Figure \ref{figure5} also indirectly depicts the effect of the secular change of the geomagnetic field upon the auroral electrojets, as observed at the GDH station.
Secular change moves the geomagnetic latitude of the GDH station roughly by 1.6$^{\circ}$ southwards and, thereby, closer to the auroral oval, from 1940s until presently.
This slow change artificially intensifies the auroral electrojets observed at GDH, as depicted by the systematic intensification of colors (positive and negative values of $\Delta Z$) in Fig. \ref{figure5}. 
However, since this secular evolution enhances both the westward and the eastward electrojet observed at GDH roughly equally, it has no significant effect on solar wind speeds extracted from the annual means of GDH $\Delta Z$ values, which are means of the 24 LT values depicted in Fig. \ref{figure5}. 


\section{HSS effect upon the H-component at sub-auroral latitudes}
\label{HSS_SOD}

Although the effect of HSSs is more purely observed on the polar side of the auroral oval, 
we have studied the long-term evolution of the WEJ intensity also using the horizontal (H) component of the Sodankyl\"a station (SOD, 67.37$^{\circ}$ geographic latitude, 26.63$^{\circ}$ geographic longitude), which is located equatorward from the auroral oval (at a sub-auroral latitude) and has one of the longest contiguous measurements since 1914. 
Contrary to GDH, the magnetic activity at  the sub-auroral SOD station is not only affected by the HSSs, but also by the storms and strong substorms due to CMEs, which expand the auroral oval more equatorwards than the HSS-related weaker substorms \citep{Holappa_Mursula_2014}. 
(Note that we can not use the Z-component to study the WEJ at sub-auroral latitudes).

In order to reduce the effect of CMEs at SOD, we have used SOD observations only around the local midnight, where the WEJ has its maximum (see Fig. \ref{figure5}) and where the connection to solar wind speed is most direct \citep{Finch_etal_2008}. 
Figure \ref{figure3_bw}b shows the annual means of $\vert \Delta H \vert$ (note that $\Delta H$ = H(d) - H(q) is negative) at SOD in the 22-02 local time sector. 
(For the earliest years 1914-1925 of SOD station we have used the station's own K indices to determine the quietest and most disturbed days.)
SOD $\vert \Delta H \vert$ values depict considerable similarity with the long-term evolution of GDH $\Delta Z$ values.
In particular, they verify that HSS activity had its long-term maximum in 1952, during the declining phase of cycle 18. 
Note that both GDH $\Delta Z$ and SOD $ \vert \Delta H \vert $ values deviate from the long-term evolution of the mid-latitude aa index (Fig. \ref{figure1_bw}b).
The differences between GDH $\Delta Z$ and SOD $ \vert \Delta H \vert$ are due to CMEs, which raise SOD $\vert \Delta H \vert$ activity in solar maximum years, especially around 1960 and 1990, above the level caused by HSSs (that are more reliably represented by GDH $\Delta Z$ values).

Despite the effect of CMEs, the annual means of SOD $\vert \Delta H \vert $ and solar wind speed are also highly correlated, and have a correlation coefficient of 0.79 which, for 45 annual data points, yields the zero correlation probability of p = 1.2$\cdot 10^{-5}$. 
The annual SW speeds derived from SOD $\vert \Delta H \vert $, together with the speeds obtained from GDH $\Delta Z$ (copied from Fig. \ref{figure4_bw}b) are shown in Fig. \ref{figure6}.
The error bars for speeds obtained from SOD $\vert \Delta H \vert $ were calculated in the same way as for GDH $\Delta Z$.
Figure \ref{figure6} shows that SOD $\vert \Delta H \vert $ values give higher speeds around the sunspot maxima of cycles 19 and 22, and raise the low level around the maximum of cycle 21 higher. 
They also yield a somewhat higher peak in 2003.
These differences are, as noted above, mainly due to the effect of CMEs, which have a larger effect to the $\vert \Delta H \vert $ values  of the sub-auroral SOD station than to GDH $\Delta Z$. 
Therefore, the results based on the GDH $\Delta Z$ more truthfully represent the long-term evolution of HSS activity than those based on SOD $\vert \Delta H \vert $.
Despite these differences, there is a good overall agreement between the results extracted from the two stations at opposite sides of the auroral electrojet.
In particular, both stations agree on the uniqueness of HSS activity in cycle 18, and that the year 1952 was the maximum year in HSS activity over the whole GMM.
Note that this agreement excludes the possibility that secular evolution would seriously modify the long-term consistency of GDH observations.

\section{Discussion and Conclusions}

We have used here a recently presented method based on the Z-component of the magnetic field at a polar cap station \citep{Lukianova_Mursula_2012} to study the occurrence of high-speed solar wind streams.
Using the observations of the longest recording polar cap (GDH) station, we have estimated the annual mean solar wind speeds over the Grand Modern Maximum (GMM) from 1920s onwards.
Since the Z-component of a polar cap station closely responds to HSSs, the method gives the annual mean speeds of the HSSs occurring during the year.
The high-speed streams having a roughly constant speed (about 800\,km/s), the annual mean speeds determined in this way reflect the relative occurrence of these streams at the Earth's orbit.


We find that persistent HSS activity occurred in the declining phase of all the 8 solar cycles studied.
During 5 out of 8 cycles (cycles 16, 19,  20,  22,  23) one year clearly stood out as the most HSS year of the respective cycle.
During two other cycles (cycles 17, 21), HSS activity was distributed among a number of consecutive years, whose HSS activity always remained below the typical level of the single-year maxima.
These results can be understood in terms of a typical evolution of coronal hole extension (solar dipole tilt activation, sometimes also called the solar excursion phase).
Such an extension dies out completely during some 7-8 rotations \citep{Mursula_Zieger_1996}, weakening its amplitude during this time and having an effective HSS occurrence time of about 5 rotations.
This structure occurs randomly in time, whence the probability of having the whole extension period within one calendar year is about 2/3, in a fair agreement with the above fraction of 5/8.
(Of course, a more refined treatment needs to take also the uneven seasonal distribution of HSSs into account).

We find that the most persistent HSS activity occurred in the early 1950s, in the declining phase of solar cycle 18.
High HSS activity continued in three successive years, with the highest activity found in the year 1952.
Considering the above statistics of HSS occurrence during the other cycles of the GMM, this period (the declining phase of cycle 18) is exceptional.
According to solar dynamo theory, the strength of solar polar magnetic fields during a sunspot minimum determines the level of sunspot activity during the next solar maximum \citep{Babcock_1961, Leighton_1964, Leighton_1969, Charbonneau_2005}. 
However, this basic tenet of solar magnetism has remained unverified even for GMM, the highest activity of measured sunspot history, because of the lack of direct observations of solar polar fields. 

The same activations of new solar flux at the mid-latitudes that lead to the polar coronal hole extensions (solar excursion phases), also, via the poleward meridional flow of the trailing flux, enhance the intensity of the recently changed new-polarity field at the poles.
Therefore, the more frequent and the more strong the activations in the declining phase are, the longer the HSSs occur at low latitudes and the stronger the solar polar fields will get.
Therefore, there is indeed a connection between the annual occurrence frequency of HSSs at the Earth's orbit in the declining phase of the solar cycle and the intensity of the solar polar fields during the subsequent minimum.
In view of this connection, our finding of exceptional high-speed stream activity in the declining phase of solar cycle 18, suggests that cycle 19 was indeed preceded by exceptionally strong polar fields during the previous sunspot minimum, thus giving interesting support for the validity of solar dynamo theory during this dramatic period of solar magnetism. 



\begin{figure}[t]
\begin{center}
\includegraphics[width=12cm]{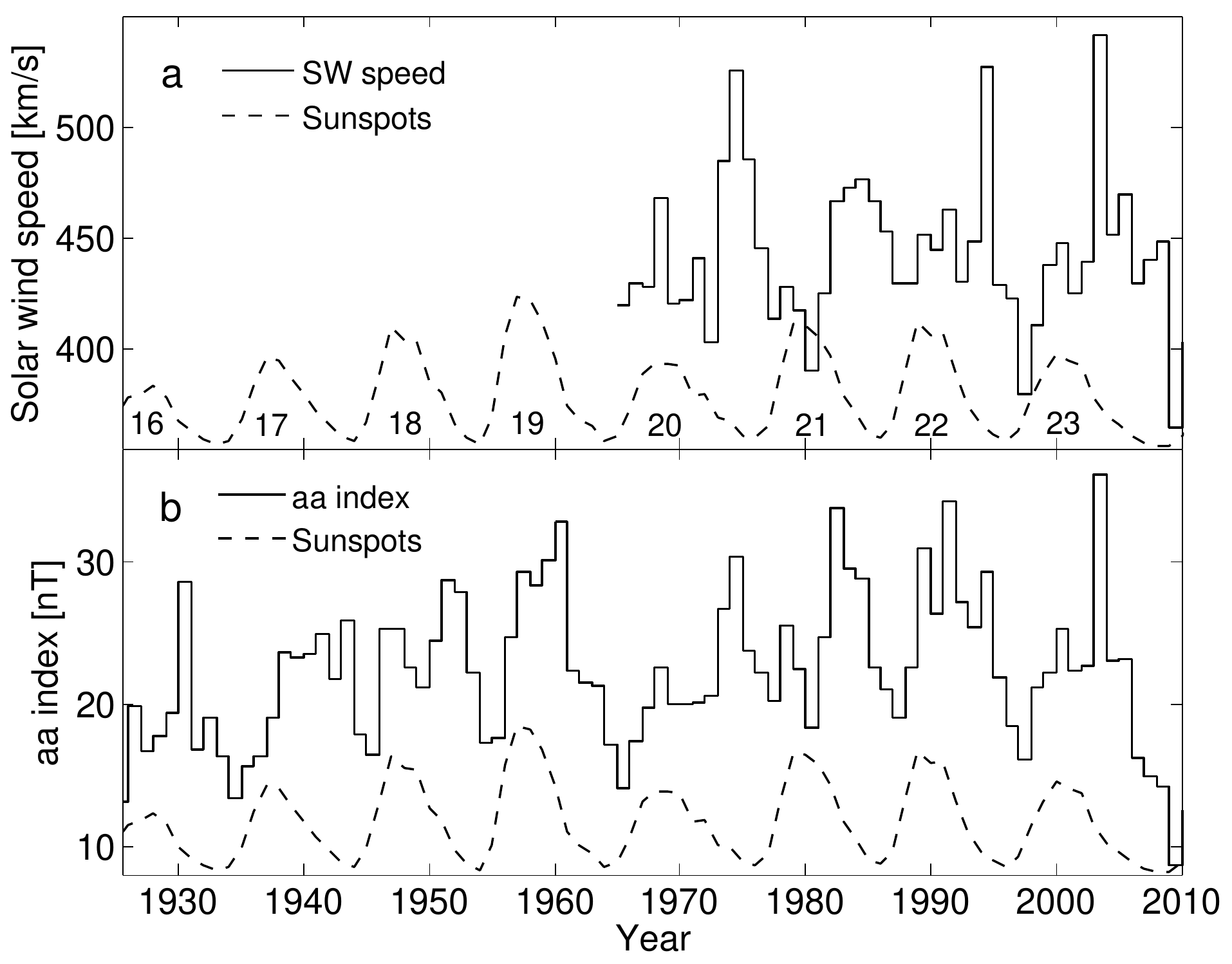}
\end{center}
\caption{Annual means of solar wind speed and the aa index. a, Satellite observed solar wind speed in 1965-2010 in units of km/s. b, aa index of geomagnetic activity in 1926-2010 in units of nT. 
Annual sunspot numbers (dashed line without scale) are included as reference in both panels. }
\label{figure1_bw}
\end{figure}

\begin{figure}[t]
\begin{center}
\includegraphics[width=12cm]{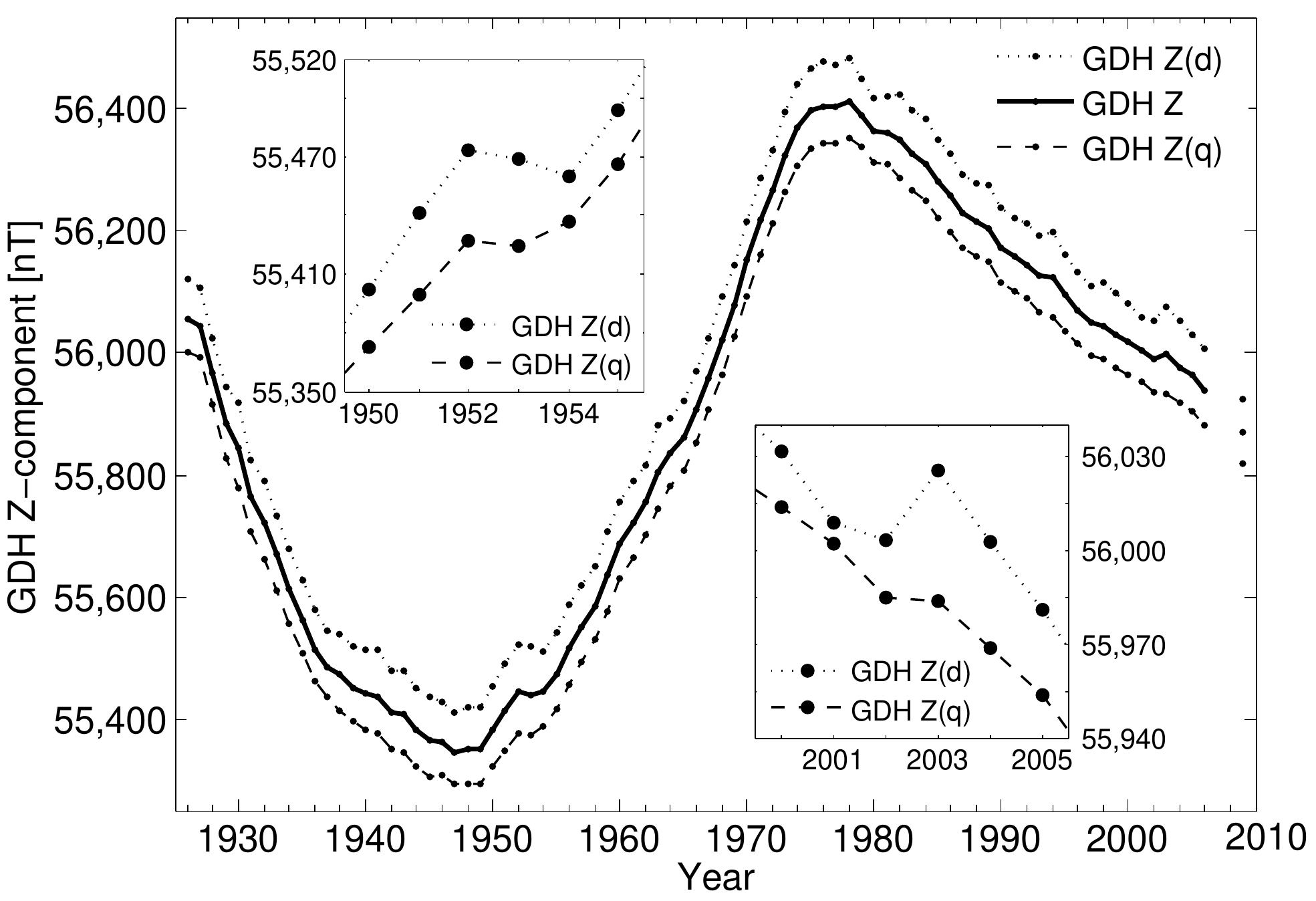}
\end{center}
\caption{Annual means of the magnetic Z-component at GDH observatory. Annual means of the Z-component are depicted separately for all days (Z; solid black line), for international quiet days only (Z(q); dashed line) and for international disturbed days only (Z(d); dotted line) in 1926-2009 in units of nT. Note the data gap in 2007-2008. Axis values are valid for the Z-curve, the dotted curve (dashed curve) has been raised (lowered, respectively) by 50 nT for better visibility. The two inserts depict the annual means for quiet and disturbed days around 1952 and 2003.  }
\label{figure2_bw}
\end{figure}

\begin{figure}[t]
\begin{center}
\includegraphics[width=12cm]{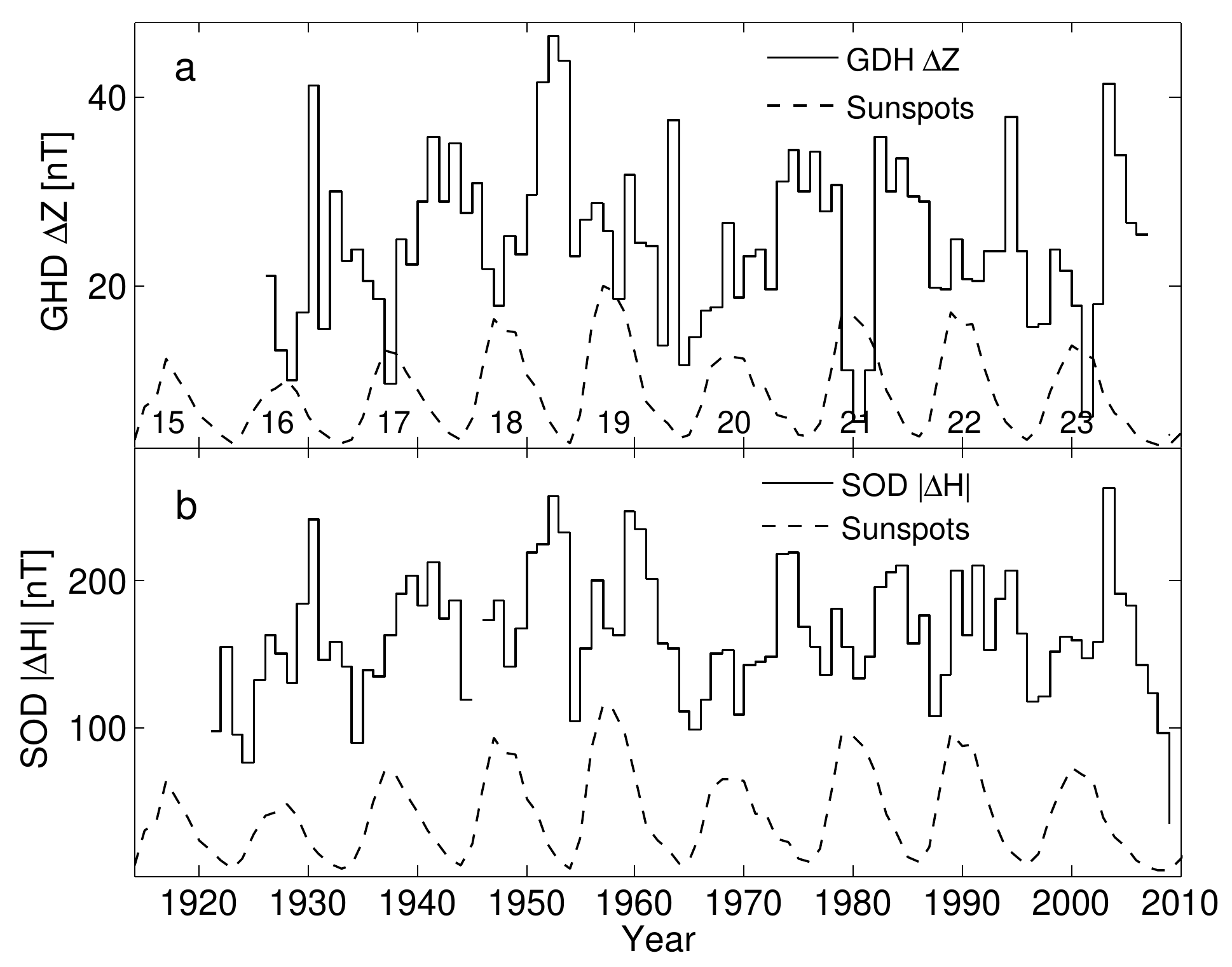}
\end{center}
\caption{Annual means of GDH $\Delta Z$ and SOD $\vert \Delta H \vert $. a, Differences $\Delta Z$ of GDH Z-component since 1926 in units of nT. b, Differences $\vert \Delta H \vert $ in SOD H-component since 1914 in units of nT. Annual sunspot numbers (dashed line without scale) are included as reference in both panels. }
\label{figure3_bw}
\end{figure}

\begin{figure}[t]
\begin{center}
\includegraphics[width=12cm]{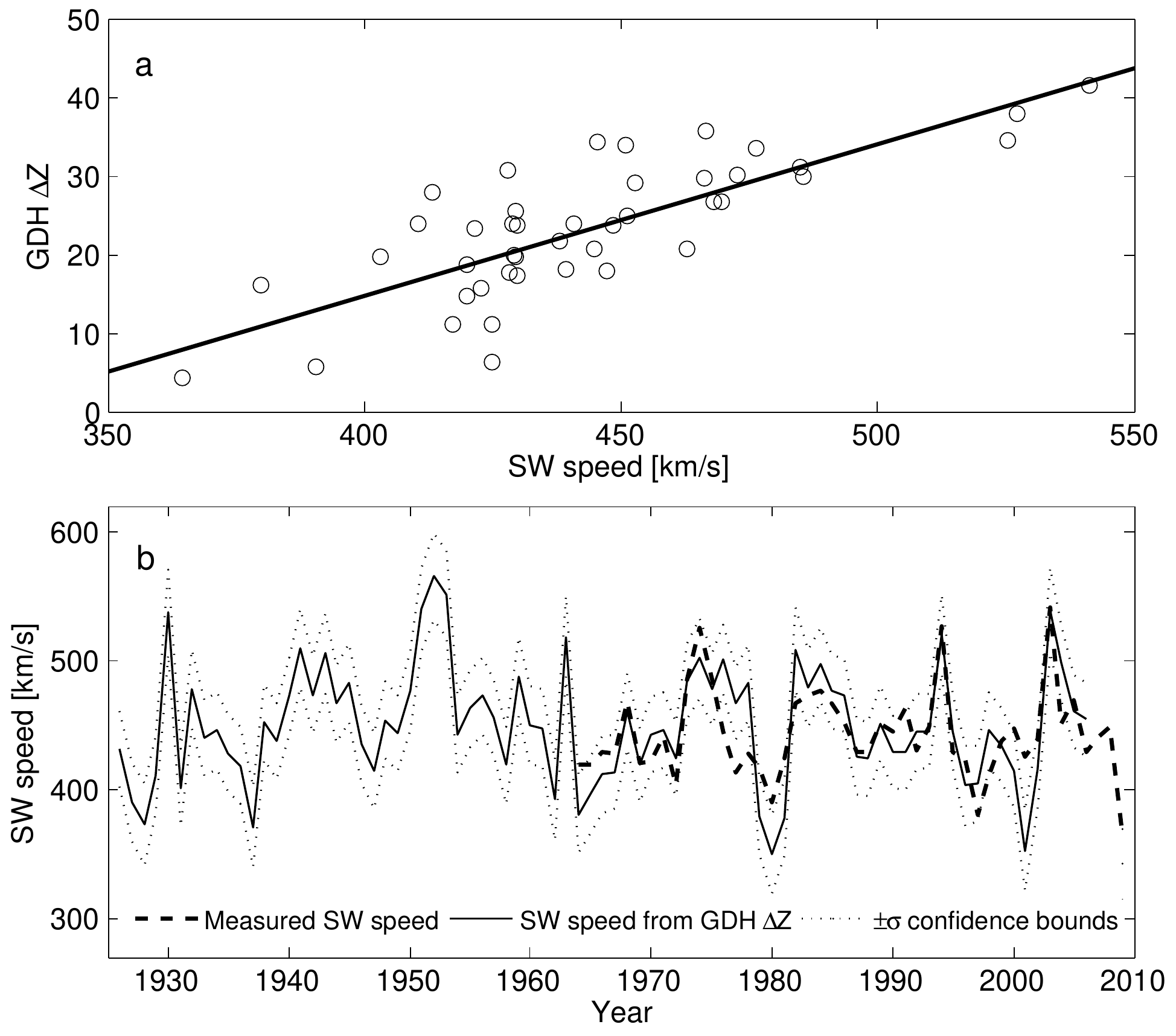}
\end{center}
\caption{a) GDH $\Delta Z$ plotted as a function of annual averages of solar wind speed. b) Annual averages of solar wind speed estimated from the GDH $\Delta Z$ (solid line) together with their 1$\sigma$ errors (two dotted lines), and observed annual averages of solar wind speed (thick dashed line). }
\label{figure4_bw}
\end{figure}

\begin{figure}[t]
\begin{center}
\includegraphics[width=12cm]{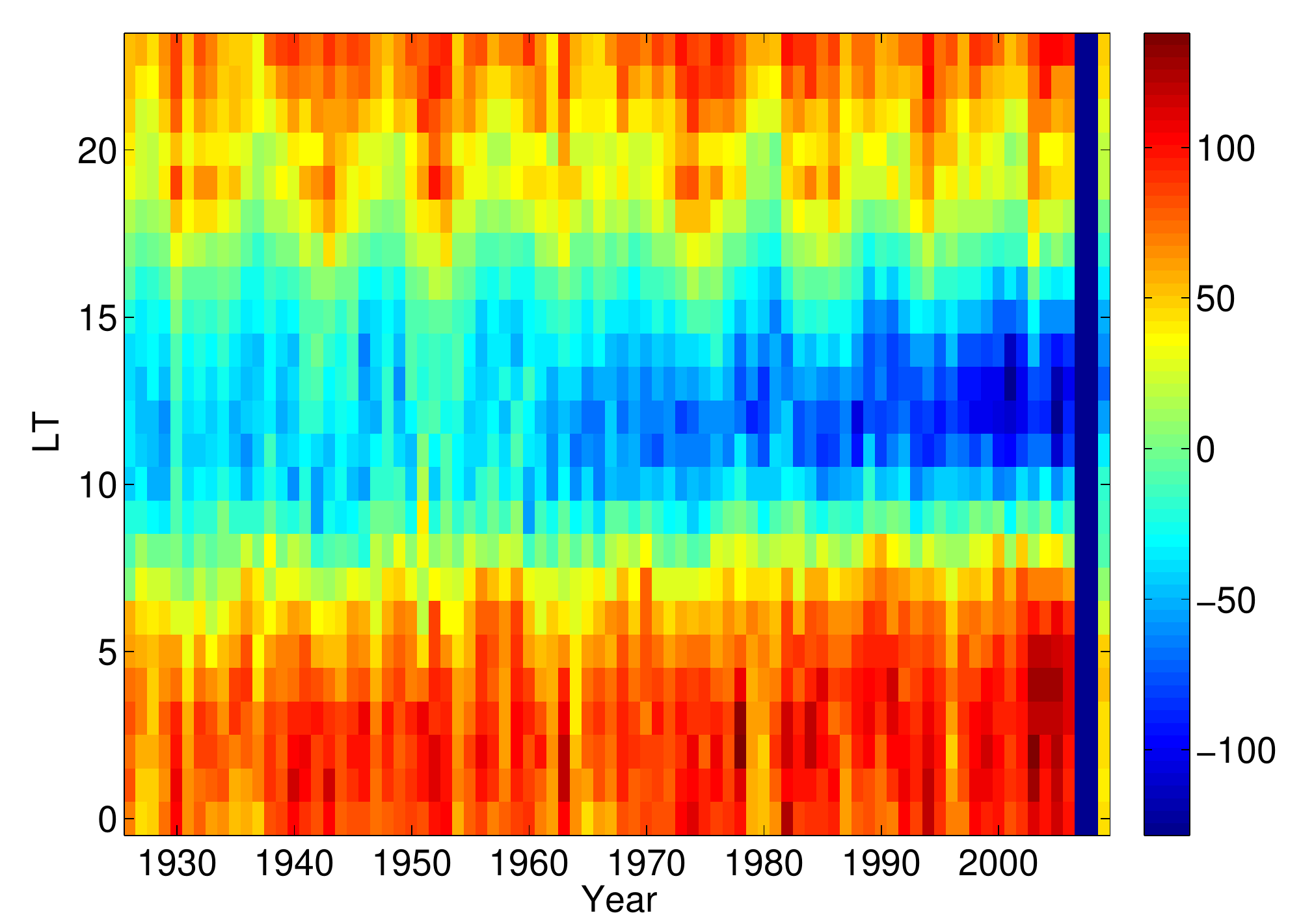}
\end{center}
\caption{Annual means of GDH $\Delta Z$ in color coding for each local time hour (vertical axis) and year (horizontal axis). For each year the annual mean of GDH $\Delta Z$ is separately calculated for each of the 24 hours of local time. Color axis gives the scale of $\Delta Z$ values in units of nT. }
\label{figure5}
\end{figure}

\begin{figure}[t]
\begin{center}
\includegraphics[width=\columnwidth]{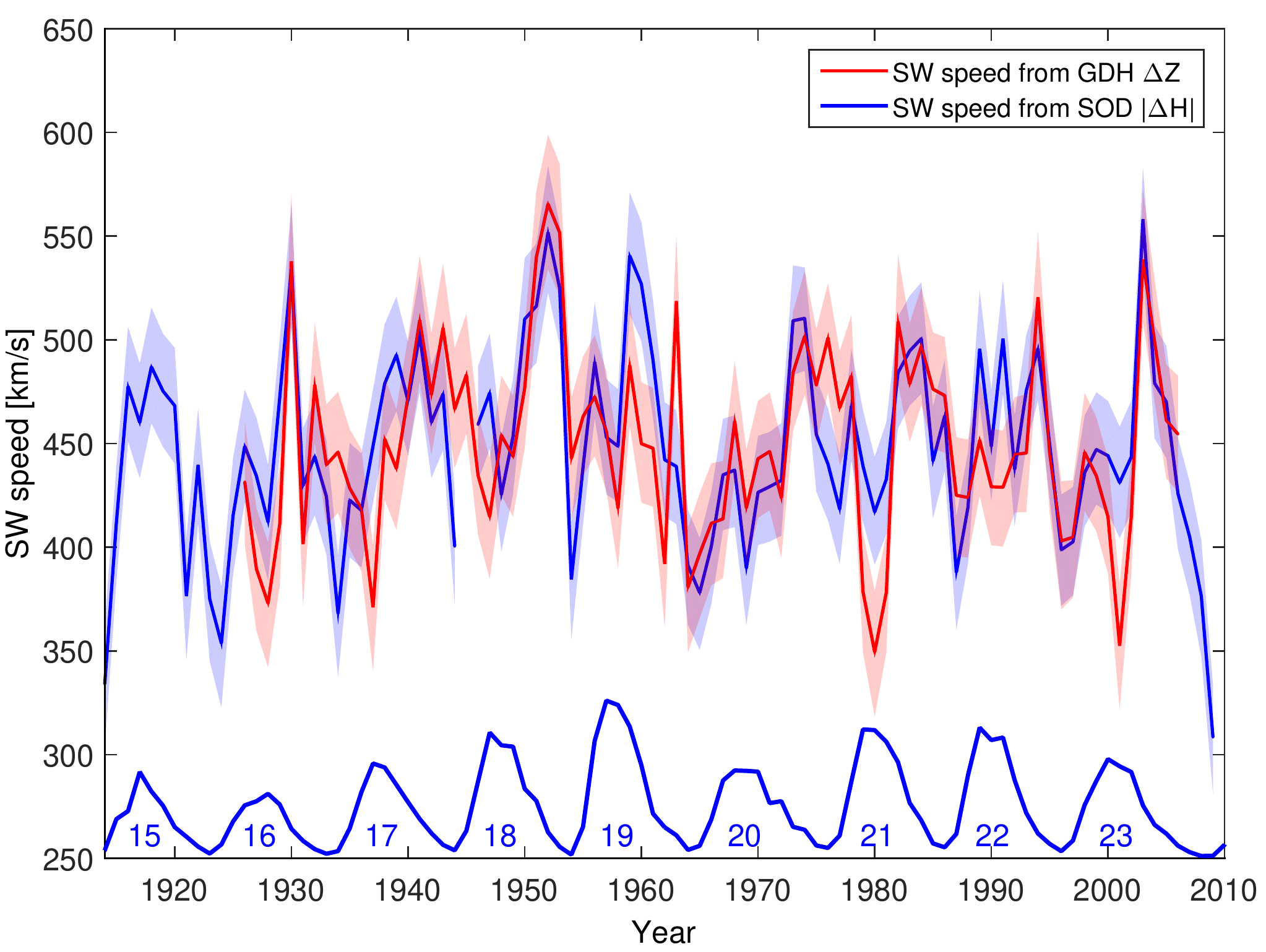}
\end{center}
\caption{Annual means of solar wind speed extracted from GDH $\Delta Z$ (red lines) and SOD $\vert \Delta H \vert $ (blue lines) with respective 1$\sigma$ errors in units of km/s denoted as shaded areas. Annual sunspot numbers (blue line without scale) are included as a  reference. }
\label{figure6}
\end{figure}

Finally, we note that some earlier studies have indicated fairly strong but no exceptional solar wind speeds in 1952 \citep{Svalgaard_Cliver_2007, Rouillard_2007, Lockwood_etal_2009, Lockwood_part4_2014}.
Typically these studies use various combinations of geomagnetic activity at mid-latitudes (like the aa index in Fig. \ref{figure1_bw}b), where the effect of CMEs is relatively larger than at auroral latitudes, especially during solar maximum years.
Finding out the detailed causes of the differences between these and the present results will, however, have to be postponed to a separate study.
Note that \citet{Lockwood_Owens_2014} have proposed a model whereby the observed centennial variation in annually averaged solar wind speed is connected to the changes in the width of the streamer belt.
While completely different in its approach, this model is conceptually in a very good agreement with the present results, excluding the detailed differences for solar wind speed in individual years, as discussed above.
We also note that, recently, a long-term proxy was developed for the hemispheric polar field strengths since 1905 using polar faculae, calibrated to the polar fiux observations during the last decades \citep{MunozJaramillo_ApJL_2013}.
They found the largest value in the northern hemisphere during cycle 18, partly supporting the present observations.

\begin{acknowledgments}

We thank all providers of the data used in this work, including the OMNI and the WDC data bases.
We acknowledge the financial support by the Academy of Finland to the ReSoLVE Center of Excellence (project no. 272157) and to projects no. 140329 and 140403.
We also acknowledge the support for this work by uniOGS of the University of Oulu and by the COST Action ES1005 "TOSCA" (\texttt{http://www.tosca-cost.eu}).

\end{acknowledgments}


\end{article}

\newpage

\end{document}